\begin{filecontents}{leer.eps}
72 31 moveto
72 342 lineto
601 342 lineto
601 31 lineto
72 31 lineto
\documentclass[epj]{svjour}
\usepackage{graphicx,psfrag}
\usepackage{dcolumn}
\usepackage{bm}
\usepackage{subfigure}
\usepackage{amsmath}
\usepackage{amssymb}

\usepackage{ulem} 
\usepackage[usenames]{color}

\begin{document}
\title{Kaon induced $\Lambda(1405)$ production on a deuteron target at DAFNE}
\author{D. Jido\inst{1} \and E. Oset\inst{2,1}
\and T. Sekihara\inst{1,3}}
%
%
\institute{
Yukawa Institute for Theoretical Physics, 
Kyoto University, Kyoto 606-8502, Japan 
\and 
Departamento de F\'{\i}sica Te\'orica and IFIC,
Centro Mixto Universidad de Valencia-CSIC,
Institutos de Investigaci\'on de Paterna, Aptdo. 22085, 46071 Valencia, Spain
\and
Department of Physics, Graduate School of Science, Kyoto University,
Kyoto, 606-8502, Japan}
\date{Received: date / Revised version: date}
%
\abstract{
The $K^{-}$ induced production of $\Lambda(1405)$ in the
$K^{-} d \to \pi \Sigma n$ reaction is investigated having in mind the conditions of the 
DAFNE facility at Frascati where kaons are obtained from the decay of slow moving $\phi$ mesons.  We find that the 
$K^{-}d \to \Lambda(1405)n$ process
favors the production of $\Lambda(1405)$ initiated by the $K^-p$ 
channel, which gives largest weight to the higher mass $\Lambda(1405)$ appearing at 1420 MeV in chiral theories.  We find that the fastest kaons from the decay of the $\phi$ are well suited to see this resonance, particularly if one selects forward going neutrons in the center of mass, which reduce the contribution of single scattering and make the double scattering dominate where the signal of the resonance appears clearer. 
} 
%


\maketitle

\section{Introduction}

The $\Lambda(1405)$ resonance is getting renewed experimental attention 
and different reactions have been used to produce it. One of the recent reactions is the  $K^- p \to \pi^0 \pi^0 \Sigma^0$ measured at $p_{K^-}$ = 514 MeV/c to 750 MeV/c in \cite{Prakhov:2004an}. COSY at Juelich has also produced it in the $p p \to K^+ p \pi^0 \Sigma^0$ reaction \cite{Zychor:2007gf}. The photoproduction of the resonance has been investigated at LEPS \cite{Niiyama:2008rt} and CLAS
\cite{Moriya:2009mx} and plans are made to continue its study in DAFNE and J-PARC among other facilities. One of the motivations to search for it in different reactions was the theoretical observation in \cite{Jido:2003cb} that
the use of chiral dynamics led to two nearby poles in the region of the 
$\Lambda(1405)$ resonance, and that different reactions gave more weight to one or the other poles, such that the shape of the cross section would change from one reaction to another.  

The $\Lambda(1405)$ has been a historical example of 
a dynamically generated resonance in 
meson-baryon coupled-channels dynamics with $S=-1$~\cite{Dalitz:1967fp}.
Modern investigations based on chiral dynamics with a unitary framework 
also reproduce well the observed spectrum of the $\Lambda(1405)$ together 
with cross sections of $K^{-}p$ to various 
channels~\cite{Kaiser:1995eg,Oset:1998it,Oller:2000fj,Oset:2001cn,GarciaRecio:2002td,Hyodo:2002pk,Hyodo:2003qa,Hyodo:2008xr}.

 The work of \cite{Jido:2003cb} also motivated further theoretical studies introducing terms of higher order in the kernel of the $K^-p$ interaction with its coupled channels \cite{Borasoy:2005ie,Oller:2006jw,Borasoy:2006sr}. The interesting thing observed in these works is that the two pole structure is always there independent of variations in the parameters allowed by the data.
 One of the states, located around 1420 MeV, couples dominantly to the $\bar KN$
channel and is very stable, while the other one, sitting around 1390 MeV and with a width of 130 MeV or larger, 
couples strongly to the $\pi \Sigma$ channel. This latter one is more model dependent but the mass is smaller than that of the other state and the width is considerably larger. Another interesting output of these works is that the results including the higher order kernel are compatible with those with the lowest order Lagrangian (the Weinberg Tomozawa term) within theoretical uncertainties \cite{Borasoy:2006sr}. Depending on the reaction mechanism and which is the channel predominantly chosen by the reaction to create the $\Lambda(1405)$, one obtains one shape or another from the superposition of the two resonances with different weights. Certainly, if one finds a reaction in which the $\Lambda(1405)$ is basically created through the entrance channel $\bar KN$, this will give more weight to the narrow, higher energy resonance, and one should see a spectrum peaking around 1420 MeV. The problem is that the $\bar KN$ threshold is already above the resonance mass. An ideal method to produce the narrow resonance is the $K^- p \to \gamma \Lambda(1405)$ reaction, since the mechanism where the photon is radiated from the initial $K^-$ is dominant. Then the emission of the photon reduces the $K^-$ energy, such as to be able to produce the  $\Lambda(1405)$, and one ensures the creation of the resonance from $K^- p$. The theoretical study for this reaction was done in \cite{Nacher:1999ni}, but the experiment is not yet done. Conversely, the study done in \cite{Hyodo:2003jw} for the $\pi^-p\to K^0\pi\Sigma$ reaction showed that this was dominated by the $\pi \Sigma$ entrance channel and this justified the peak seen in the $\pi \Sigma$ invariant mass at around 1400 MeV in the experiment \cite{Thomas:1973uh}. Other reactions like the  $p p \to K^+ p \pi^0 \Sigma^0$ \cite{Zychor:2007gf}, require a mixture of both resonances as seen in  \cite{Geng:2007vm}.
 One of the strongest support for the idea of the two $\Lambda(1405)$ states comes from the experiment of \cite{Prakhov:2004an}, $K^- p \to \pi^0 \pi^0 \Sigma^0$, and the subsequent theoretical analysis of the reaction done in \cite{magas}. In this case one of the $\pi^0$ is ``radiated" from the initial proton, the $K^- p$ system loses energy, and as a consequence the $\Lambda(1405)$ production is initiated by the $K^- p$ system, exciting the higher mass state that couples mostly to the the $K^- p$ system and producing a peak in the $\pi \Sigma$ spectrum around 1420 MeV. 
 
 Another great support for the idea of the two $\Lambda(1405)$ states comes from the $K^{-} d \to \pi\Sigma n$ reaction \cite{Braun:1977wd} with kaons in flight, where a clear peak is seen in the $\pi \Sigma$ spectrum around 1420 MeV. This reaction was studied recently in \cite{Jido:2009jf} and it was found that the rescattering mechanism of the $K^-$, in which the $K^-$ scatters on the neutron, loses some energy, propagates to the proton and produces the $\Lambda(1405)$ on the proton target, dominates the reaction in this invariant mass region. Since  the production of the $\Lambda(1405)$ is done from the $K^- p$ channel, it also excites predominantly the higher mass $\Lambda(1405)$ state, which shows up in the 
$\pi \Sigma$ spectrum with a peak around 1420 MeV. 

The detail of kaons in flight is important. Indeed, in \cite{Jido:2009jf} it was shown that, together with the double scattering, where the $\Lambda(1405)$ shows up cleanly, there is the contribution of the single scattering, the impulse approximation. This contribution peaks around a value of the $\pi\Sigma$ invariant mass, corresponding to the invariant mass of the kaon in flight and one nucleon at rest. If one has sufficient kaon kinetic energy this peak and the one of the 
 $\Lambda(1405)$ are quite separated. If one goes at threshold with kaons at rest, the single scattering peak is around $m_K +M_N$, does not have resonant structure and by interference with the double scattering distorts the shape of the  $\Lambda(1405)$ contribution, and the reaction is not suited to investigate the properties of this resonance.  In \cite{Esmaili:2009rf} a different opinion is expressed and claims are made that the impulse approximation can provide the shape of the resonance. We argue below that this is not the case. Indeed, one can get $K^- p $ invariant masses below threshold for kaons at rest, only through the Fermi motion of the deuteron. Yet, to reach invariant masses of the order of 1400 MeV, down from the threshold of 1432 MeV, one needs momentum components of the order of 300 MeV/c. Obviously, our knowledge of the deuteron wave function for such large momentum components is extremely poor. In fact, standard wave functions of the deuteron have dropped by two orders of magnitude at 300 MeV/c, with appreciable differences between different models. This is four orders of magnitude in the cross section, where the uncertainties are certainly not smaller than one order of magnitude. Even the concept of wave function, when  one has a transfer of energy of 32 MeV to a second nucleon, is ambiguous. Trying to reconstruct a shape of the resonance, by dividing an experimental cross section by the theoretical deuteron wave function squared has very large uncertainties and should not be used. The possible use of an ``experimental" wave function for these high momentum components should be equally discouraged, since processes involving such momentum transfers have contributions from two body mechanisms difficult to quantify. 
 
   DAFNE produces intense kaon beams from $\phi$ decay. Recently the $\phi$ mesons can even be produced with a small momentum of 12 MeV/c, by means of which the kaons have a range of 80 MeV/c to 130 MeV/c when they reach the target. We shall take advantage of this and investigate the production process for the most energetic kaons, looking at different kinematical conditions that maximize the clean production of the resonance.

\section{Formulation}

\label{sec:form}

We briefly sketch here the formalism of \cite{Jido:2009jf} for this reaction. Further details can be seen in that paper.

\begin{figure}
\centerline{\includegraphics[width=6.5cm]{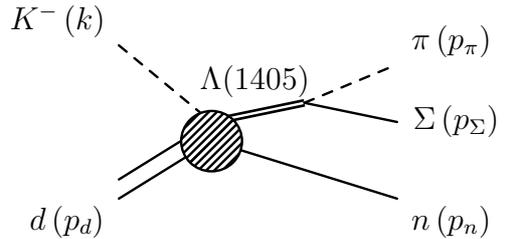}}
\caption{Kinematics of the $K^{-} d \to \pi \Sigma n$. \label{fig1}}
\end{figure}

We consider $\Lambda(1405)$ production induced by $K^{-}$ with a deuteron
target, $K^{-} d \rightarrow \Lambda(1405) n$. 
The $\Lambda(1405)$ produced in this reaction decays into $\pi \Sigma$ with $I=0$
as shown in fig.~\ref{fig1}. The $\Lambda(1405)$ is identified by the $\pi \Sigma$ invariant mass spectra of this reaction. 
Figure~\ref{fig1} also gives the kinematical variables of
the initial and final particles. The kinematics of the three-body final state is completely
fixed by five variables, the $\pi\Sigma$ invariant mass $M_{\pi\Sigma}$, the neutron 
solid angle $\Omega_{n}$ in the c.m.\ frame and the pion solid angle $\Omega_{\pi}^{*}$
in the rest frame of $\pi$ and $\Sigma$~\cite{Amsler:2008zz}. 
Thus the differential cross section of this reaction can be written as
\begin{equation}
d \sigma = \frac{1}{ (2\pi)^{5}} 
\frac{M_{d}M_{\Sigma} M_{n}}{4k_{\rm c.m.}E_{\rm c.m.}^{2}}
   \, |{\cal T}|^{2}
   |\vec p_{\pi}^{\, *}|\,
   |\vec p_{n}|\, dM_{\pi\Sigma} d\Omega_{\pi}^{\,*} d\Omega_{n}
   \label{eq:difcross}
\end{equation}
where $\cal T$ is the $T$-matrix of this reaction, 
$E_{\rm c.m.}$ is the center of mass energy, $k_{c.m.}$ is the kaon CM momentum
and
$\vec p_{\pi}^{\, *} $ is the pion momentum in the rest frame of $\pi$ and $\Sigma$.
The pion momentum $|\vec p_{\pi}^{\, *}|$ in the $\pi \Sigma$ rest frame 
can be fixed by the invariant mass $M_{\pi\Sigma}$.

\begin{figure}
\begin{center}
\centerline{\includegraphics[width=8.5cm]{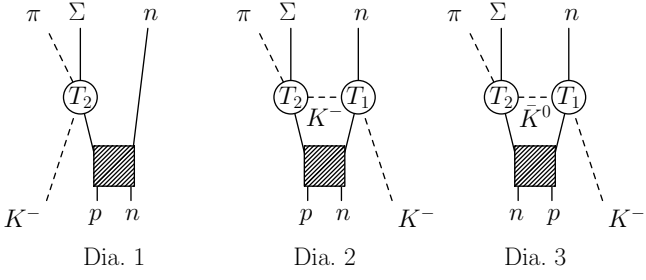}}
\caption{Diagrams for the calculation of the $K^{-}d \to \pi\Sigma n$ reaction.
$T_{1}$ and $T_{2}$ denote the scattering amplitudes for $\bar KN \to \bar KN$
and $\bar K N \to \pi \Sigma$, respectively.  \label{fig2}}
\end{center}
\end{figure}

The $\Lambda(1405)$ production is investigated by 
limiting the kinematics with the invariant mass of the final $\pi \Sigma$ state
around 1350 to 1450 MeV, in which the resonating $\pi\Sigma$ forms the 
$\Lambda(1405)$
and the resonance contribution may dominate the cross section.  

In this $\Lambda(1405)$ dominance
approximation, we have three diagrams for this reaction as shown in fig.~\ref{fig2}.
The left diagram of fig.~\ref{fig2} expresses  the $\Lambda(1405)$ production 
in the impulse approximation. We refer to this diagram as direct production process.  The middle and right diagrams are for two-step
processes with $\bar K$ exchange. We refer to these diagrams 
as double scattering diagrams. 

In fig.~\ref{fig2}, $T_{1}$ and $T_{2}$ denote $s$-wave scattering amplitudes of 
$\bar K N \to \bar KN$ and $\bar K N \to \pi \Sigma$, respectively. These amplitudes 
are calculated in a coupled-channel approach based on chiral dynamics. 
%

The ${\cal T}$  matrix appearing in Eq. (\ref{eq:difcross}) is given by the sum of the contribution of the three diagrams of  fig.~\ref{fig2}

\begin{equation}
   {\cal T} =  {\cal T}_{1} + {\cal T}_{2} + {\cal T}_{3}
\end{equation}
where the different amplitudes are given by 

\begin{equation}
  {\cal T}_{1} = T_{K^{-}p \rightarrow \pi \Sigma}(M_{\pi\Sigma}) \, \varphi(\vec p_{n} - \frac{\vec p_{d}}{2}). \label{eq:T1}
\end{equation}

\begin{eqnarray}
  {\cal T}_{2}& =&  T_{K^{-}p \rightarrow \pi \Sigma}(M_{\pi\Sigma}) 
 \int \frac{d^{3}q}{(2\pi)^{3}} \frac{\tilde \varphi (\vec q+\vec p_{n}-\vec k - \frac{\vec p_{d}}{2})}{q^{2}-m_{K}^{2} + i\epsilon}
   \nonumber \\ && \times
 T_{K^{-}n \rightarrow K^{-}n}(W_{1}) \ . \label{eq:T2}
\end{eqnarray}

\begin{eqnarray}
  {\cal T}_{3}& =& - T_{\bar K^{0}n \rightarrow \pi \Sigma}(M_{\pi\Sigma}) 
  \int \frac{d^{3}q}{(2\pi)^{3}} \frac{\tilde \varphi (\vec q+\vec p_{n}-\vec k - \frac{\vec p_{d}}{2})}{q^{2}-m_{K}^{2} + i\epsilon}
  \nonumber \\ && \times
  T_{K^{-}p \rightarrow \bar K^{0}n}(W_{1})  \ . \label{eq:T3}
\end{eqnarray}
with $\varphi(\vec p_{n} - \frac{\vec p_{d}}{2})$ the deuteron wave function in momentum space and 
\begin{eqnarray}
   q^{0} &=& M_{N} + k^{0} - p_{n}^{0}\ , \\
   W_{1} &=& \sqrt{(q^{0}+p_{n}^{0})^{2}-(\vec q + \vec p_{n})^{2}} \ .
\end{eqnarray}
For $q^0$ we have assumed that the deuteron at rest has energy 
$2M_N -B$, and we have taken half of it for one nucleon, 
neglecting the small binding energy. The variable $W_1$ depends, 
however, on the running $\vec{q}$ variable.

It should be noted, as it was also the case in \cite{Jido:2009jf}, 
that taking an average value of $\vec q$,  we can fix $W_{1}$ by the 
external variables as
\begin{equation}
   W_{1}  = \sqrt{(M_{N}+k^{0})^{2} - \vec k^{\, 2}},
   \label{eq:W}
\end{equation}
which allows us to take the $T_{\bar KN \to \bar KN}$ amplitude 
out of the loop integral and 
produces results practically equal to those obtained with the values of these variables depending on the loop variable. 
Note also that $T_{\bar KN \to \pi\Sigma}$, which is the amplitude producing 
the resonance shape, has its argument fixed in all cases by the external  
$\pi \Sigma$.

The $T$-matrix for the diagram 1 given in fig.~\ref{fig2} has been calculated 
in the impulse approximation in which the incident $K^{-}$ and the proton in 
the deuteron produce the $\pi \Sigma$ and the neutron behaves 
as a spectator of the reaction. As for the external particles, we have taken
 the wave functions of the incident kaon and the particles in the final state as plane waves.

The double scattering requires a loop integral over the $\vec q$ variable in 
Eqs.~(\ref{eq:T2}) and (\ref{eq:T3}) with the deuteron wave function. 
Although the procedure is standard in multiple collision in nuclei, we find it appropriate to 
sketch the connection to a conventional loop evaluation with Feynman diagrams. 
The loop integral would be
\begin{equation}
    \int \frac{d^{3} q}{(2\pi)^{3}} \frac{1}{E_{\rm tot} - k^{0} -H_{0NN}} t_{NN} \phi_{NN}
    \frac{T_{\bar KN \to \bar KN} T_{\bar KN \to \pi\Sigma}}{q^{2} - m_{K}^{2} + i \epsilon} \ ,
\end{equation}
where $H_{0}$ is the free $NN$ Hamiltonian, $t_{NN}$ the appropriate $NN$ scattering
matrix and $\phi_{NN}$ the free $NN$ wave function. By using the definition of 
the scattering matrix 
\begin{equation}
   t_{NN} \phi_{NN} = V_{NN} \Psi_{NN}
\end{equation}
where $V_{NN}$ is the $NN$ interaction potential and 
$\Psi_{NN}$ is the actual $NN$ wave function, the integral becomes 
\begin{equation}
  \int \frac{d^{3} q}{(2\pi)^{3}} \frac{1}{E_{NN}  -H_{0NN}} V_{NN} \Psi_{NN}
    \frac{T_{\bar KN \to \bar KN} T_{\bar KN \to \pi\Sigma}}{q^{2} - m_{K}^{2} + i \epsilon} \ .
\end{equation}
Now taking into account the Schr\"odinger equation 
\begin{equation}
  \Psi_{NN} = \frac{1}{E_{NN}-H_{0NN}} V_{NN} \Psi_{NN}  \ ,
\end{equation}
one rewrites the integral as 
\begin{equation}
    \int \frac{d^{3} q}{(2\pi)^{3}} \Psi_{NN} \frac{T_{\bar KN \to \bar KN} T_{\bar KN \to \pi\Sigma}}{q^{2} - m_{K}^{2} + i \epsilon}
\end{equation}
which is the integral that appears in Eqs.~(\ref{eq:T2}) and (\ref{eq:T3}).

\section{Results}

 In fig.~\ref{fig3} we show $d \sigma/dM_{\pi \Sigma}$ for $\pi^+ \Sigma^-$, $\pi^- \Sigma^+$ and $\pi^0 \Sigma^0$ production for 130 MeV/c $K^{-}$ momenta in the lab.~frame. What we see is a clear dominance of the single scattering.
 However, we observe a certain structure, more visible in the $\pi^+ \Sigma^-$ spectrum, with a peak 
at the invariant mass squared, $M_{\pi\Sigma}^{2}=(k+p_{N})^{2}$, with $p_{N}$ at rest in the deuteron rest frame, where the deuteron wave function has its largest weight. This peak has to do with the deuteron wave function and not with the amplitude $T_{K^{-}p\to \pi\Sigma}$ of Eq.~(\ref{eq:T1}). Then we see
another peak in  the region of 1420~MeV. 
This second peak comes from the shape of the $T_{K^{-}p\to \pi\Sigma}$ amplitude of Eq.~(\ref{eq:T1}), which reflects the excitation of the higher energy $\Lambda(1405)$ resonance appearing around 1420~MeV.
We also see that the mechanism of double scattering also peaks around 1420 MeV and there is interference with the single scattering amplitude which, nevertheless, leaves the peak at about the same position. Yet, since the shape of the double scattering is cleaner and does not have the double shoulder, it would be interesting to find some situation in which it becomes dominant. It is easy to find such kinematical set up, since in the impulse approximation the neutron is a spectator and goes backwards in the $K^- d$ CM frame. If we demand that the neutron goes forward we shall be reducing drastically the single scattering contribution, giving room to the double scattering one. The results of imposing angular cuts can be seen in fig.~\ref{fig4}, 
in which $\theta_{\rm c.m.}^{n}$ is the angle 
between the outgoing neutron and the incident $K^{-}$ in the CM frame.  
We can see that for large angles of the neutron the impulse approximation is largely dominant, yet, as soon as we make the angles of the neutron smaller, the contribution of the single scattering becomes gradually smaller and at angles smaller than 30 degrees the double scattering has become dominant, with the total cross section showing a clean peak around 1420~MeV. To make it more precise, in fig.~\ref{fig5} we show the quantity $d \sigma/dM_{\pi \Sigma}\,d\cos(\theta_{\rm c.m.}^{n})$ for different angles and we find that for angles below 32 degrees the double scattering contribution is dominant. Hence, for practical purposes, and in order to gain statistics, gathering all events with angles smaller than 32 degrees in the CM is ideal in order to show a clean contribution of the $\Lambda(1405)$. The cross sections of 0.1 mb/MeV, from Fig. \ref{fig4}, are large enough to be observed in the DAFNE set up. A larger span of angles would also show a clear peak, with gain in statistics, since even at 60 degrees the peak is visible, as one can appreciate in figs.~\ref{fig4} and~\ref{fig5}.
 
\begin{figure}
\begin{center}
\centerline{\includegraphics[width=7cm]{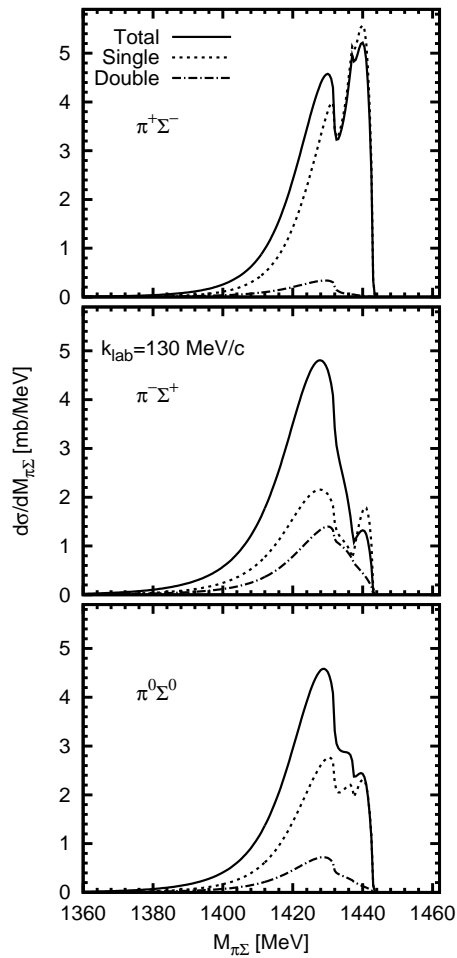}}
\caption{$\pi\Sigma$ invariant mass spectra of $K^{-}d \to \pi \Sigma n$
with 130 MeV/c of $K^{-}$ incident momentum. The three panels correspond to 
$\pi\Sigma$ spectra for different $\pi\Sigma$ charge combinations,
$\pi^{+}\Sigma^{-}$ (upper panel), $\pi^{-}\Sigma^{+}$ (middle panel)
and $\pi^{0} \Sigma^{0}$ (lower panel). In each panel, the solid line denotes 
the total contributions of the three diagrams, while the dotted and 
dash-dotted lines show the calculations 
from the single and double
scatterings, respectively. 
}
\label{fig3}
\end{center}
\end{figure}

\begin{figure}
\begin{center}
\centerline{\includegraphics[width=8.5cm]{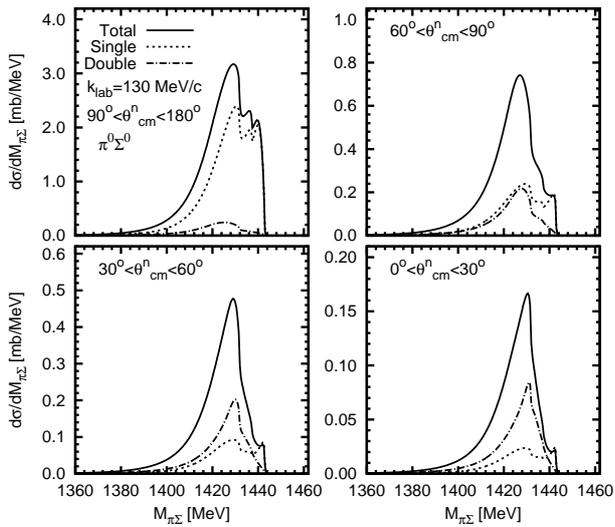}}
\caption{$\pi\Sigma$ invariant mass spectra of $K^{-}d \to \pi^{0}\Sigma^{0}n$
with 130 MeV/c of incident $K^{-}$ momentum imposing angular cuts 
for the emitted neutron with respect to the incident $K^{-}$ in the CM frame, 
$90^{\circ}<\theta^{n}_{\rm c.m.}<180^{\circ}$ (up-left),
$60^{\circ}<\theta^{n}_{\rm c.m.}<90^{\circ}$ (up-right),
$30^{\circ}<\theta^{n}_{\rm c.m.}<60^{\circ}$ (down-left) and
$0^{\circ}<\theta^{n}_{\rm c.m.}<30^{\circ}$ (down-right).
In each panel, the solid line denotes 
the total contributions of the three diagrams, while the dotted and 
dash-dotted lines show the calculations 
from the single and double
scatterings, respectively. 
}
\label{fig4}
\end{center}
\end{figure}

\begin{figure}
\begin{center}
\centerline{\includegraphics[width=8.5cm]{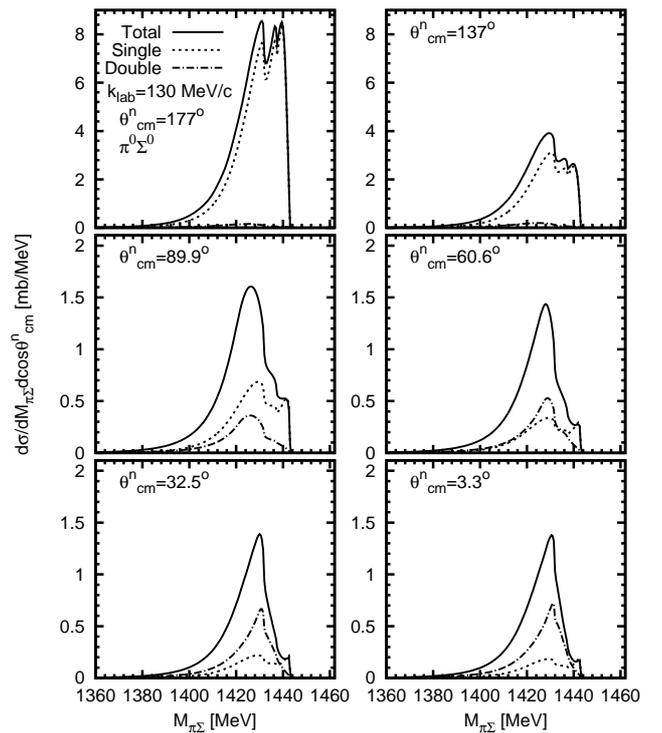}}
\caption{Differential $\pi\Sigma$ invariant mass spectra of $K^{-}d \to \pi^{0}\Sigma^{0}n$
for 130 MeV/c of incident $K^{-}$ momentum fixing the angle 
between the emitted neutron and the incident $K^{-}$ in the CM frame.  
In each panel, the solid line denotes 
the total contributions of the three diagrams, while the dotted and 
dash-dotted lines show the calculations 
from the single and double
scatterings, respectively. 
}
\label{fig5}
\end{center}
\end{figure}

By producing the $\phi$ with a finite small momentum in the new DAFNE set up, one obtains kaons with larger momentum than before, when the $\phi$ was produced at rest. In what follows, we show the results for two other kaon momenta, within the present kaon momentum range, to emphasize the value of taking the kaons with larger momentum. 
In fig.~\ref{fig6}, we show the results equivalent to fig.~\ref{fig5} for momenta of the kaons 100 MeV/c and 80 MeV/c.  At $k= 100$ MeV/c we can see in fig.~\ref{fig6} that the contribution of the single scattering dominates the cross section at neutron backward angles, and one finds two peaks, one coming from the resonance shape and the other one from the threshold effect. The two peaks are closer among themselves than in the case of  $k= 130$ MeV/c. The final shape should then not be confused with the shape of the resonance. If one goes to neutron forward angles one can see that the single scattering is still as large as the double scattering one, and it shows its two peak structure, such that the shape of the global distribution is not the shape of the resonance.  This momentum is not good to extract the properties of the $\Lambda(1405)$ resonance. The situation becomes worse for $k= 80$ MeV/c, as one can see in fig.~\ref{fig7}. Here, even at forward angles, the single scattering dominates the cross section and the shape that comes out does not reflect the shape of the resonance. Fortunately, the situation for  $k=$ 130 MeV/c is satisfactory, with a clear dominance of the double scattering process and still a visible separation of the resonance and threshold peak contributions of the single scattering, such that the resulting peak around 1420 MeV clearly reflects the shape of the upper energy $\Lambda(1405)$ state peaking at 1420 MeV. 
 Hence, the small increase in the kaon momentum from the former situation, where the kaons came from $\phi$ decay at rest,
is most welcome and helps to get a cleaner signal of the  $\Lambda(1405)$ resonance. 

 One may wonder that, since double scattering becomes more important than single scattering in the kinematic conditions chosen, the triple scattering could also be important. However, as a general rule for inclusive processes, the single scattering is
 more important than the double and this one more important than the triple. We have observed this here, since the strength of the peak of single scattering is much bigger than that of double scattering. Yet, the
 single scattering has a very limited kinematics, a very restricted phase space,
 but this is not longer the case for double and triple scattering. Then the
 selection of forward neutron angles forces the single scattering out of its phase
 space, except for the finite momentum components of the deuteron, but does not
 put constraints on double scattering nor triple scattering, such that the
 triple scattering will remain small compared to the double.
   Note that there is an important difference between the case of the break up process that we have here and the coherent scattering of $K^-$ with deuteron without deuteron breakup. In the latter case the two nucleons in the deuteron stick together and it is relatively easy for the kaons to rescatter. This is indeed the case as shown in \cite{Toker:1981zh,Barrett:1999cw,kamalov}. But in the inclusive process, after the first scattering the nucleons move apart making more difficult the rescattering of the kaons with this pair of nucleons.

We also do not consider double scattering diagrams
with pion exchange in which the $\pi$  and the $\Sigma$ are emitted
separately in the 
$T_{1}$ and $T_{2}$ amplitudes given in fig.~\ref{fig2},
respectively. It was discussed in \cite{Jido:2009jf} that
such diagrams would give a smooth background in the $\pi \Sigma$
invariant mass spectra.

\begin{figure}
\begin{center}
\centerline{\includegraphics[width=8.5cm]{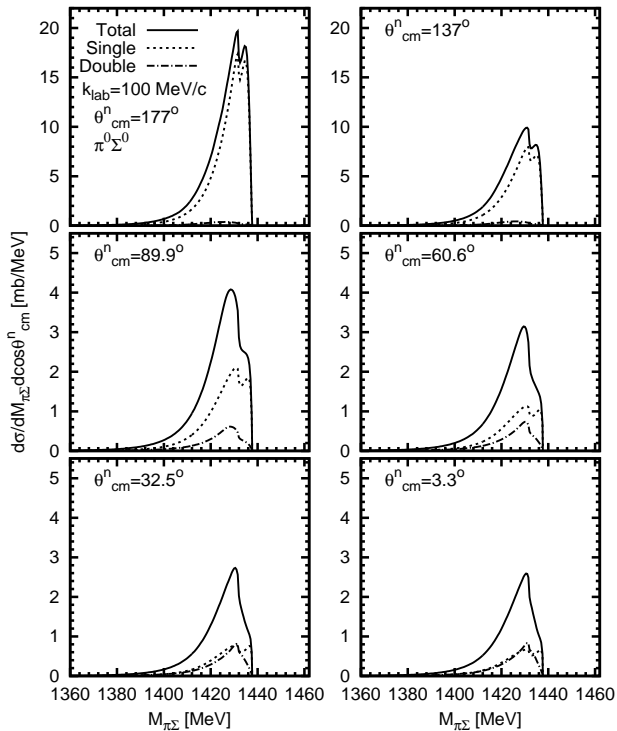}}
\caption{Differential $\pi\Sigma$ invariant mass spectra of $K^{-}d \to \pi^{0}\Sigma^{0}n$
for 100 MeV/c of incident $K^{-}$ momentum fixing the angle
between the emitted neutron and the incident $K^{-}$ in the CM frame.  
In each panel, the solid line denotes 
the total contributions of the three diagrams, while the dotted and 
dash-dotted lines show the calculations
from the single and double
scatterings, respectively.  
}
\label{fig6}
\end{center}
\end{figure}

\begin{figure}
\begin{center}
\centerline{\includegraphics[width=8.5cm]{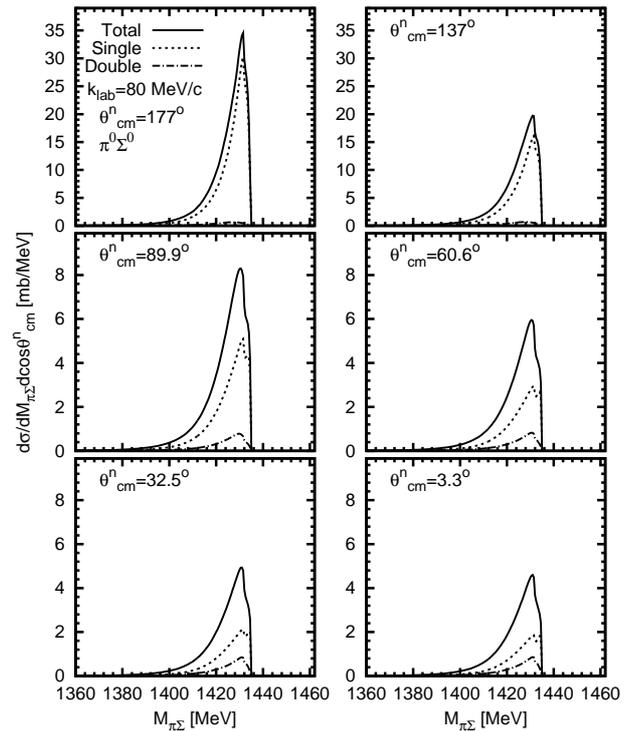}}
\caption{Differential $\pi\Sigma$ invariant mass spectra of $K^{-}d \to \pi^{0}\Sigma^{0}n$
for 80 MeV/c of incident $K^{-}$ momentum fixing the angle between the emitted neutron and the incident $K^{-}$ in the CM frame.
In each panel, the solid line denotes 
the total contributions of the three diagrams, while the dotted and 
dash-dotted lines show the calculations 
from the single and double
scatterings, respectively. 
}
\label{fig7}
\end{center}
\end{figure}

\section{Summary}
\label{sec:summary}

We have studied the $K^{-}$ induced production of $\Lambda(1405)$
with a deuteron target in the reaction $K^{-} d \to \pi \Sigma n$, with the conditions of the DAFNE set up, where kaons are produced from the decay of slow moving $\phi$ mesons. The process proved well suited to investigate the predicted $\Lambda(1405)$ state around 1420 MeV, which couples mostly to $\bar{K}N$, since in the $K^{-} d \to \pi \Sigma n$ process,
the $\Lambda(1405)$ resonance is produced by the $\bar KN$ channel.

 We found that for the most energetic kaons coming from the decay of the moving $\phi$ mesons, both the single scattering and the double scattering gave rise to a peak in the region of 1420 MeV for the spectrum of $\pi \Sigma $ invariant masses. Yet, it was also found that the contribution of single scattering can be easily distorted by a peak around threshold of the $K^-p$ channel. In order to obtain a cleaner signal, we evaluated the cross section at forward angles of the neutron and we found that the contribution of single scattering is drastically reduced and that of the double scattering takes over, leading to a cleaner signal of the $\Lambda(1405)$ resonance. The general rule is that having kaons in flight, at moderate energies of the kaons, becomes advantageous to distinguish the  $\Lambda(1405)$ signal, since this automatically reduces the contribution of the single scattering versus the double scattering one. This is not so because the contribution of the single scattering is smaller, but because it peaks at a different region of invariant masses. Yet, we could see that, given the energy constraints at DAFNE, it was still possible to drastically reduce the contribution of single scattering at the expense of looking at a kinematical region in the phase space, with neutrons forward, where the single scattering contribution is reduced but the double scattering, which has more flexibility on momentum sharing, is not so much affected.
 
  In view of the results obtained here we can encourage to do this experiment at DAFNE where the fluxes of kaons are large enough to make the experiment feasible. Indeed, a first estimate \cite{nevio} indicates that with the energy resolution for charged particles of 2-4 \% of the kinetic energy and 4.2 MeV for neutrons for the energies of the experiment, measured with 20-35 \% efficiency in the KLOE calorimeter (not necessarily needed if the $\pi$ and $\Sigma$ are detected), the shape of the distribution can be obtained with the needed resolution. Taking into account present fluxes and the calculated cross sections one would need about eight months of beam time \cite{nevio}.

\section*{Acknowledgements}
%
One of us, E.~O.~wishes to acknowledge useful discussions with Nevio Grion and Alessandra Filippi. This work was partly supported by
the Grant-in-Aid for Scientific Research
from MEXT and JSPS (Nos.
   22105507,    	
   22740161,    	
   20540273         
   and 22-3389),   
the collaboration agreement between the JSPS of Japan and the CSIC of Spain,  
and the Grant-in-Aid for the Global COE Program 
``The Next Generation of Physics, Spun from Universality and Emergence" 
from MEXT of Japan.
This work was also supported in part by DGICYT contract number
FIS2006-03438. 
We acknowledge the support of the European Community-Research
Infrastructure Integrating Activity
``Study of Strongly Interacting Matter" 
(acronym HadronPhysics2, Grant Agreement n. 227431) under 
the Seventh Framework Programme of EU.
One of us, T.~S.~acknowledges the support by the Grand-in-Aid for JSPS fellows. 
This work was supported in part by DFG (SFB/TR 16, ``Subnuclear Structure of Matter'').
This work was done under
the Yukawa International Program for Quark-hadron Sciences (YIPQS).

%
%

\end{document}